\documentclass[review]{elsarticle}

\usepackage{hyperref}

\journal{Journal of \LaTeX\ Templates}

\usepackage{xcolor,colortbl}
\newcommand{\Rafa}[1]{\textcolor{blue}{\small [#1 -- Rafa]}}
\newcommand{\Felipe}[1]{\textcolor{blue}{\small [#1 -- Felipe]}}
\usepackage{multirow}
\usepackage{amssymb}
\usepackage{amsmath}
\usepackage{lineno,hyperref}
\usepackage{booktabs}

\begin{document}

\begin{frontmatter}

\title{Classification of Emotions and Evaluation of Customer Satisfaction from Speech in Real World Acoustic Environments} 

\author{Luis Felipe Parra-Gallego$^{a,b}$, Juan Rafael Orozco-Arroyave$^{a,c}$\corref{mycorrespondingauthor}}

\cortext[mycorrespondingauthor]{Corresponding author}
\ead{lfelipe.parra@udea.edu.co, rafael.orozco@udea.edu.co}

\address[mymainaddress]{GITA Lab. Faculty of Engineering, University of Antioquia UdeA, Medell\'in, Colombia.}
\address[mysecondaryaddress]{Konecta Group S.A.S. Medell\'in, Colombia.}
\address[mysecondaryaddress]{Pattern Recognition Lab. Friedrich Alexander University, Erlangen-Nuremberg}

\begin{abstract}
This paper focuses on finding suitable features to robustly recognize emotions and
evaluate customer satisfaction from speech in real acoustic scenarios. 
The classification of emotions is based on standard and well-known corpora and
the evaluation of customer satisfaction is based on recordings of real opinions given
by customers about the received service during phone calls with call-center agents. 
The feature sets considered in this study include two speaker models, 
namely x-vectors and i-vectors, and also the well known feature set introduced in
the Interspeech 2010 Paralinguistics Challenge (I2010PC). Additionally, we
introduce the use of phonation, articulation and prosody features extracted 
with the DisVoice framework as alternative feature sets to robustly model 
emotions and customer satisfaction from speech.
The results indicate that the I2010PC feature set is the best approach to classify
emotions in the standard databases typically used in the literature.
When considering the recordings collected in the call-center, without any control
over the acoustic conditions, the best results are obtained with our articulation 
features. 
The I2010PC feature set includes 1584 measures while the articulation
approach only includes 488 measures. We think that the proposed approach 
is more suitable for real-world applications where the acoustic conditions are not
controlled and also it is potentially more convenient for industrial applications.

\end{abstract}


\end{frontmatter}

\section{Introduction}

A common practice in call-centers consists of recording 
and storing customer calls for posterior analyses. 
Those calls are listened to and evaluated with the aim of 
improving the quality of service (QoS). 
This procedure is usually hand made by randomly taking 
small samples from the total set of calls. During the QoS 
evaluation several aspects are rated including 
whether the call-center agent resolved the customer's 
problem or need, whether the agent's service was efficient
and timely, whether the tone and volume of the agent's voice 
was not offensive, whether the answer was provided calmly, 
whether customer got angry, among others~\cite{zweig2006automated}.
This analysis procedure has two main disadvantages: 
(1) There is a double cost for rating the calls: answering 
the calls and evaluating them; and 
(2) only a very little number of call among the total set is
evaluated. These aspects limit the effectiveness in the QoS 
evaluation and improvement process~\cite{zweig2006automated}. 

Automatic speech emotion recognition (SER) systems were 
introduced several years ago and have evolved a lot since then.
The main drawback in this topic relies on the fact that there are
almost no databases with real (non-acted) emotions appropriately
rated by experts. Even though this main limitation, it has been
shown that SER systems are suitable to help call-center managers in monitoring 
and optimizing the QoS provided by their agents~\cite{mishne2005automatic}. 
These systems can potentially detect the emotional state of agents and/or 
customers and hence provide a QoS index. Abnormal changes 
in service patterns like increasing number of angry customers 
can be detected~\cite{paralinguistics2013}. 
SER systems can efficiently and timely evaluate the total incoming calls. 
In spite of the relevance of these systems to assure and
improve QoS, the research community is still working on the way to find
better corpora appropriately labeled and systems that are robust against 
non-controlled acoustic conditions, which are the real-world scenarios 
in which the technology is actually useful.

The aim of this work is to evaluate different approaches typically used 
when developing automatic SER systems, and also to introduce a novel approach
with features that focus on modeling phonation, articulation and prosody aspects
that may vary when the emotional state of the speaker is altered.
The approaches are tested upon well-known corpora typically used to evaluate
automatic SER systems and also upon a new corpus that consists of
recordings where customers give their opinion about the service provided 
by call-center agents. 
The recordings are labeled by experts in QoS according to whether the 
service requested by the customer was satisfactorily provided or not. 
The main differences
between the typical speech emotional databases and the one introduced here
include: (1) The first group of corpora includes labels about 
specific emotions produced by the speakers, while the call-center corpus
only includes labels about customer satisfaction, and 
(2) most of the emotional speech databases consists of acted emotions and are 
recorded under controlled-acoustic conditions, while the call-center database comprises 
recordings of real conversations of customers
and service agents that are collected without any control over the recording
process or the channel. Additionally, these conversations are
labeled by experts in customer service, which is the real way of evaluating 
these kinds of interactions in real industrial applications.

An exhaustive search of feature extraction techniques was carried out 
to find the most effective ones to detect emotional states in different 
acoustic environments. Since different investigations in SER have shown 
that x-vector, i-vector, and I2010PC features can efficiently represent 
emotions from speech~\cite{eyben2015geneva,guo2019exploration,xia2012using,jalal2019learning,khaki2015continuous,pappagari2020x}, we considered those sets 
in this study along with the articulation, phonation, and prosody 
features extracted with the DisVoice framework~\cite{orozco2018neurospeech}. 
The classification of emotions is performed with a support vector machine (SVM) 
classifier, which is a standard method, well-known and has shown to 
be the among the most robust methods in different
scenarios\cite{wang2015speech,eyben2015geneva,semwal2017}.

In order for our results to be comparable with other studies, three 
``standard'' databases are considered for the classification of emotions in speech: 
(1) Interactive Emotional Dyadic Motion Capture (IEMOCAP), 
(2) Emotional database (EMODB), and 
(3) Ryerson Audio-Visual Database of Emotional Speech and Song (RAVDESS). 
To evaluate the robustness of the system to accurately evaluate customer 
satisfaction in real acoustic conditions, experiments with a call-center 
database, namely KONECTADB are included. 

The rest of the paper is distributed as follows: 
Section~\ref{sec:state-of-the-art} presents 
an overview of related works. 
Section~\ref{sec:contributions} presents the main contributions.
Section~\ref{sec:dataset} describes the databases considered
in this study.
Section~\ref{sec:methodology} introduces details of the methodology 
followed in this work. Section~\ref{sec:results} presents the experiments and 
results, and finally, Section~\ref{sec:conclusions} includes the conclusions 
derived from the study and future work.

\section{Related works~\label{sec:state-of-the-art}}
Many techniques have been studied to develop SER systems.
In~\cite{wang2015speech}, the authors proposed to recognize different 
emotions included in the German database EMODB (happiness, 
boredom, neural, sadness, anger, and anxiety).
The recordings were modelled with a set of 120 harmonic features 
along with their $\Delta$ and $\Delta \Delta$. 
Their minimum, maximum, mean, median, and standard deviation were 
also computed at an utterance level, producing 1800-dimensional 
feature vectors per sample. Speaker-independent multi-class
classification was performed using SVMs and the authors reported
average recall values of 92.\%, 71.48\%, 87.46\%, 91.42\%, 98,29\%, 
and 91.92\% for the aforementioned emotions, respectively.
For the multi-class classification scenario, the authors report
an average accuracy of 79.51\%, which is a relatively high and 
optimistic accuracy because the validation strategy reported by 
the authors was a k-fold cross-validation, which does not guarantee 
unbiased results.
The authors in~\cite{semwal2017} worked also with the EMODB corpus
and modeled the emotions with temporal, spectral and cepstral
features. Statistical functionals were computed per feature vector 
at an utterance level. The authors reported a maximum accuracy of
80\% in the multi-class classification of the different emotions
included in the dataset. Note that all of the experiments were 
speaker-dependent, which leads to optimistic and biased results.
Another relevant work in the topic of automatic emotion recognition
was presented in~\cite{eyben2015geneva}. 
The authors proposed a feature set that included frequency, 
energy/amplitude, and spectral parameters to model the speech signals. 
Mean, standard deviation, 20th, 50th and 80th percentiles, 
range between 20th and 80th percentile were the functionals 
computed per feature vector. 
The authors used a wide variate of datasets (TUM AVIC, GEMEP, EMODB, 
SING, FAU AIBO, and Vera-am-Mittag) to map information from
the affective speech domain to the binary arousal and valence
representation.
An SVM classifier was trained following a leave-one-speaker-out 
cross-validation strategy to classify between high vs. low arousal 
and between positive vs. negative valence. The accuracy reported in the first 
scenario was 79.71\% and 66.44\% for the second one.
Other approaches typically used to create SER systems are based on
speaker representation models. For instance,
the authors in~\cite{xia2012using} used i-vectors to classify 
emotions in two different datasets: (1) USC AudioVisual
data~\cite{busso2004analysis} and (2) IEMOCAP~\cite{iemocap},
which are acted and spontaneous, respectively. 
The authors trained an SVM for the classification experiments and
compared their approach w.r.t. the feature set introduced in the
INTERSPEECH 2010 Paralinguistic Challenge~\cite{schuller2010interspeech}.
According to their results, the system based on i-vectors yielded better 
performance (91.1\% for USC and 71.3\% for IEMOCAP). 
It is important to highlight that these results were achieved on
speaker-dependent experiments and the hyper-parameter optimization
procedure was not explained in detail, so it is not possible to known whether
these results are optimistic and possibly biased. 

More robust SER systems have been developed using deep learning 
techniques in the recent years. 
In~\cite{fayek2017evaluating}, the authors compared 3 different 
network architectures: 
(1) Convolutional Neural Network (CNN); 
(2) Artificial Neural Network (ANN); and 
(3) Long Short Term Memory (LSTM). 
The IEMOCAP dataset was used to evaluate the approaches following an
eight-fold-leave-two-speakers-out cross-validation scheme in 
all experiments. 
Log-Fourier transform -based filter-bank with coefficients distributed
on the Mel scale were extracted. Because the authors performed a
multi-class classification at a frame level, they assumed that 
frames belonging to a given utterance convey the same emotion as the 
parent utterance. Silence class was added using the labels generated 
by a voice activity detection system. 
The method was compared with prior works in the literature related 
to multi-class emotion classification that used the same database. 
Given that most works reported results at an utterance level, 
the posterior class probabilities computed for each frame in 
an utterance were averaged across all frames and an utterance-based 
label was selected based on the maximum average class probabilities. 
The authors reported that their system outperformed all the 
other methods in up to 2\% of accuracy.
In~\cite{pappagari2020x}, the authors explored possible dependencies
between speaker recognition and emotion recognition topics. 
They first applied the transfer learning technique to transfer information
from a Resnet-based pre-trained speaker-identity-based model to 
an emotion classification task. 
They also explored how the performance of a speaker recognition 
model is affected by different emotions. 
The authors evaluated their experiments on three different
datasets: IEMOCAP, MSP-Podcast, and Crema-D. 
Two approaches were explored: x-vectors extraction and 
replacement of the speaker-discriminative output layer with 
an emotion classification layer and then fine-tune the hyper-parameters. 
A total of 23 MFCCs with 25ms frame-size and 10ms frame-shift were
extracted. The model was validated on a speaker-independent scenario. 
The best results were obtained with the transfer learning approach with
accuracies of up to 70.3\%, 58.46\%, and 81.84 for IEMOCAP, 
MSP-Podcast, and Crema-D, respectively. 
In the same year, the authors in~\cite{issa2020speech}
proposed a one-dimensional CNN architecture to model 
emotions from speech. The input to the architecture was 
based on MFCCs, chromagram, Mel-scale spectrogram, Tonnetz 
representation, and spectral contrast features extracted 
from the audio files. 
The authors performed speaker-independent experiments with
4 classes of the IEMOCAP corpus and reported accuracies 64.3\%.
The experiments with RAVDESS and EMODB were speaker-dependent and
the reported accuracies were 71.61\% and 81.1\%, respectively. 
Also in 2020, a novel technique was introduced 
in~\cite{sajjad2020clustering}. The authors proposed a method where
key segments are selected based on a radial basis function network 
(RBFN) similarity measurement. The segments were grouped following 
the k-means algorithm as follows: The audios were divided 
into multiple chunks of 500ms and the RBFN similarity was computed.
If consecutive segments did not exceed a given threshold, the 
number of clusters was increased by one. Thus, the number of 
clusters $k$ dynamically changed. Once the segments were 
clustered, the nearest segment with respect to the centroid was 
selected for each cluster to generate a new sequence, which 
was then converted into a spectrogram. The spectrogram was passed 
through a CNN network (pre-trained ResNet101) to extract 
high-level features. These features were normalized and used as inputs 
to a deep bi-directional long short term memory (BiLSTM), 
which made the final decision about which emotions were present
in a given recording. 
The authors considered the three ``standard'' databases IEMOCAP, 
EMODB, and RAVDESS and reported speaker-independent 
accuracies of 72.25\%, 85,57\%, and 77.02\% for IEMOCAP, 
EMODB, and RAVDESS, respectively. 

From the literature review presented above it can be observed that
most of the studies in SER classify emotions using the 
EMODB, IEMOCAP, and RAVDESS databases.
Although these are considered as standard in the topic of SER, there 
is still a gap between the development of these systems and their application
in real-world scenarios, where acoustic conditions are not controlled and
the emotions are not acted.

\section{Contributions of this work\label{sec:contributions}}

With the aim to address the problem of automatic SER and also the problem
of modeling customer satisfaction from speech recordings collected under 
non-controlled conditions, this study introduces the use of phonation,
articulation and prosody features extracted with the DisVoice framework.
The feature sets are also used in different emotional speech corpora
typically considered in the literature. The results obtained with the introduced
feature sets are compared with respect to three different approaches: 
two speaker models namely, i-vectors and x-vectors, and the 
Interspeech 2010 Paralinguistics Challenge (I2010PC) feature
set~\cite{schuller2010interspeech}.
The results show that our approach is competitive when considering the 
``standard'' emotional speech databases and it is the better one when considering
the recordings with the opinions of customers of the call-center, which were 
collected under non-controlled acoustic conditions.

\section{Data~\label{sec:dataset}}
This work considers three speech emotional databases commonly used in 
the literature of SER: (1) IEMOCAP, (2) RAVDESS, and (3) EMODB. 
Each corpus contains audio recordings with emotional content.
They constitute the standard databases for the training and 
evaluation of SER models. 
Besides these well-known corpora, we introduce here the 
KONECTADB corpus, which was created with audio recordings
of a call-center. This corpus was collected by Konecta Group
S.A.S in Medellín, Colombia an it is used in this work to 
evaluate the proposed approach in real-world acoustic conditions. 
All datasets are down-sampled to 8kHz. Further details of each
corpus are presented below.

\subsection{IEMOCAP\label{subsec:iemocap}}
This is an audio-visual database that consists of approximately 
12 hours of recordings including video, speech, motion capture 
of the face, and the transliterations corresponding to a total of 
10039 recordings~\cite{iemocap}. 
The audios were originally sampled at 16kHz with 16-bit resolution. 
The database is divided into five recording sessions. 
Two actors (one male and one female) performed scripted and 
improvised scenes. The database was annotated by multiple 
annotators into four emotional labels: anger, happiness, 
sadness, neutral, and frustration. There are about 10000 samples
per class and the annotations are based on the average of the
labels assigned by the four labelers.

\subsection{RAVDESS}
This is a multimodal database with emotional speech and
songs~\cite{livingstone2018ryerson}. 
Speech recordings of 24 actors (12 male and 12 female) are included.
Each actor produced two lexically-matched statements 
in neutral north American English accent. Seven expressions with different
emotional content were produced by the actors: calm, happy, sad, 
angry, fearful, surprise, and disgust. 
Besides one expression produced with neutral emotional content,
the other expressions were produced twice, with normal and 
strong level of emotional intensity.  
There is a total of 1440 recordings sampled at 16kHz with 16-bit
resolution. Each class contains 192 samples except the neutral
one with only 96 samples.

\subsection{EMODB}
This database contains recordings of 10 German actors (5 male and 5 
female) who produced 10
utterances~\cite{burkhardt2005database}. 
Seven emotions are labeled in the recordings: anger, boredom, 
disgust, anxiety, happiness, sadness, and neutral.
The recording process was performed in ideal acoustic conditions
and using a professional audio setting. The distribution of
samples among the emotions is not even, i.e., there are emotions
with much less recordings than others.

\subsection{\label{ref:konecta-database}KONECTADB}
\begin{figure*}[!h]
\centering
   \includegraphics[width=\textwidth]{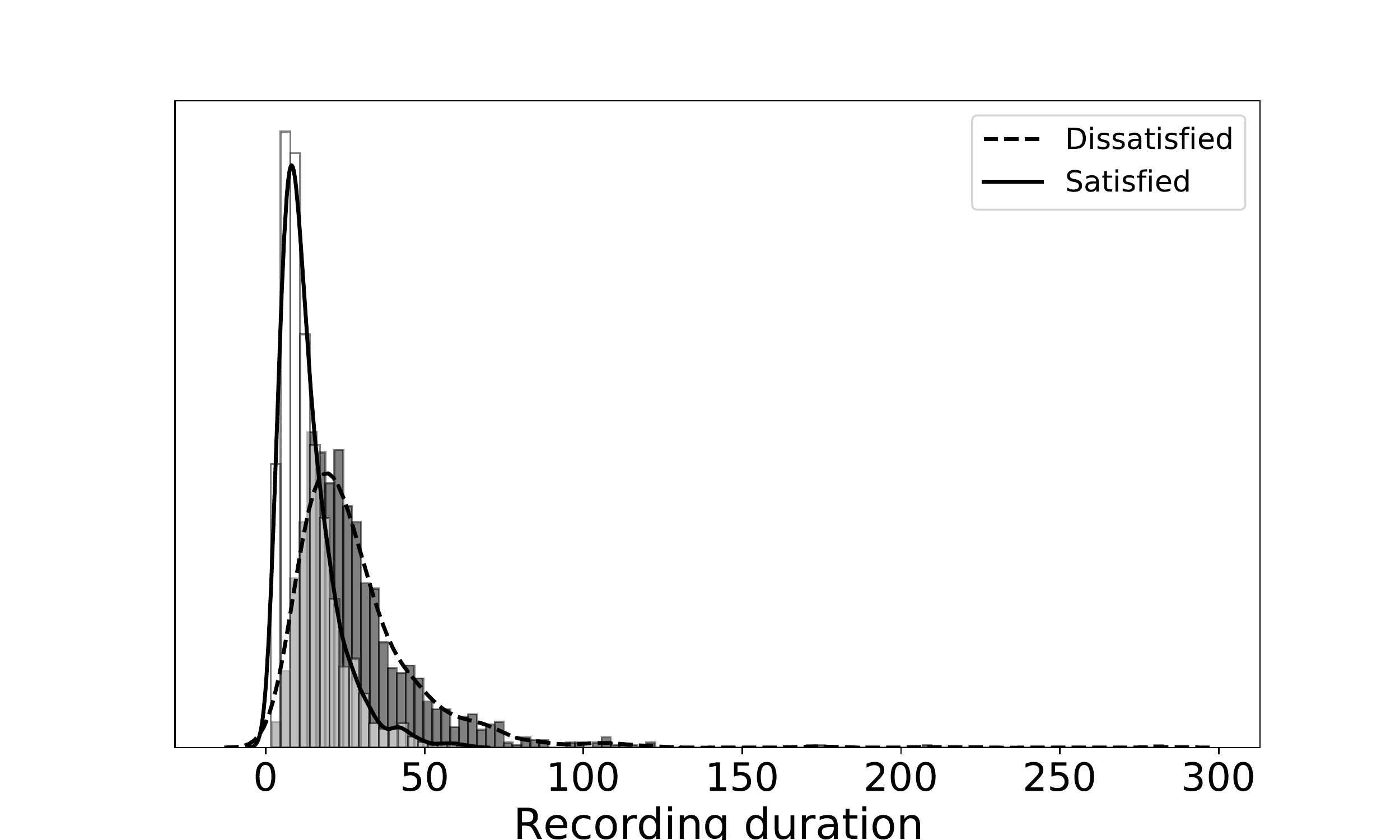}
\hfill
\caption{Distribution of the duration for satisfied and dissatisfied classes in the KONECTADB.}
\label{fig:dur_distribution}
\end{figure*}

This database contains voicemails which are recorded at 
the end of phone-calls between customers and service-agents.
In those voicemails the customers give spontaneous evaluations
of the service provided by the agent. 
The customers were informed that their speech was going to be 
recorded. Due to the nature of the service, it is assumed that 
the speakers in these recordings are all adults. 
The audios were recorded at a sampling frequency of $8$~kHz 
and 16-bit resolution. The corpus contains 2364 recordings 
annotated by experts in QoS (i.e., the labelers listed to each 
audio file and evaluated whether the customer was satisfied or not). 
The experts labeled the audios as satisfied, dissatisfied, 
or neutral according to the degree of satisfaction 
expressed by the customer in the voice-mail. 
We only considered satisfied and dissatisfied categories in this study.
There is a significant difference in the length of satisfied and 
dissatisfied recordings ($t$-test with $p \ll 0.05$).
This can also be observed in Figure~\ref{fig:dur_distribution}.
Note that the dissatisfied class contains longer recordings and 
higher length variability than the other class. 
Besides, gender-balance is validated through a chi-square test ($p\approx1$). 
Additional information of the database is provided in
Table~\ref{table:database}. 

\begin{table}[ht!]
\centering
\caption{Data distribution in KONECTADB.}
\label{table:database}
\begin{tabular}{lcc}
\toprule
\rowcolor[HTML]{FFFFFF} 
\multicolumn{1}{c}{\cellcolor[HTML]{FFFFFF}{\color[HTML]{000000} \textbf{}}} & \textbf{Dissatisfied} & \textbf{Satisfied} \\ \hline
\rowcolor[HTML]{FFFFFF} 
{\color[HTML]{000000} Number of samples}                                      & 1259              & 1105              \\
\rowcolor[HTML]{FFFFFF} 
Duration ($\mu\pm\sigma$)   & 34$\pm$23\,s    & 16 $\pm$ 11\,s    \\
\rowcolor[HTML]{FFFFFF} 
Number of male                                                                & 711               & 532               \\
Number of female                                                              & 548               & 573               \\ \bottomrule
\end{tabular}
\end{table}

\section{Methodology~\label{sec:methodology}}
\begin{figure*}[!h]
\centering
   \centering
   \includegraphics[width=\textwidth]{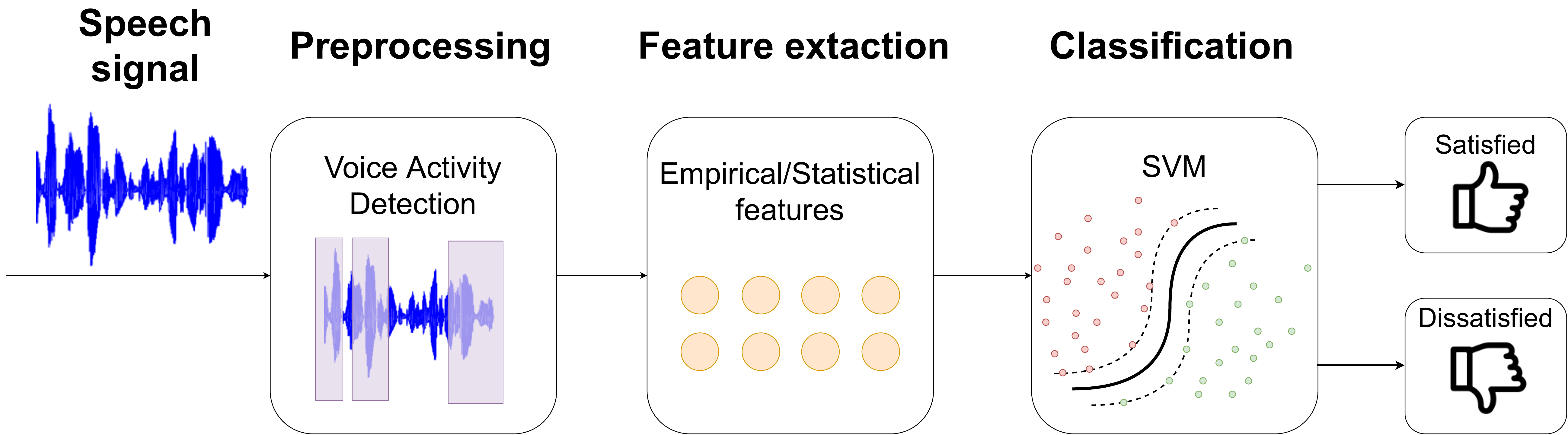}
\hfill
\caption{Methodology.}
\label{fig:methodology}
\end{figure*}

Figure~\ref{fig:methodology} illustrates the overall methodology followed
in this work which includes pre-processing, feature extraction, and
classification.

\subsection{Preprocessing\label{sec:preprocessing}}

Voice activity detection (VAD) was applied to the recordings of the 
KONECTADB to remove long silence segments. The VAD algorithm consists of a 
pre-trained model based on time-delay neural networks. It 
was trained using the Kaldi toolkit\footnote{\url{http://kaldi-asr.org/models.html}}~\cite{Povey_ASRU2011}.

\subsection{Feature extraction\label{sec:features}}
Three different approaches are considered in this study: two speaker models,
the I2010PC feature set and the feature sets extracted with DisVoice, which
include phonation, articulation, and prosody measures. Each feature set is
described below.

\subsubsection{i-vectors}
Earlier studies on speaker recognition methods to model speaker traits 
through high-dimensional super-vectors were based on 
Gaussian Mixture Models (GMMs) adapted from an Universal 
Background Model (UBM)~\cite{campbell2006support,campbell2006svm,you2008svm}. 
That approach produces a big set of parameters, which requires a 
large dataset to be trained. To solve this problem, the authors
in~\cite{dehak2009support} proposed a low-dimensional vector
representation called identity vector (i-vector). 
The low-dimensional space is defined by a 
matrix called the total variability matrix $T$, which models 
both speaker and channel variability. The new model is
represented as follows:

\begin{equation}
    \label{eqn:factor-i-vector}M = m + Tw
\end{equation}
where $M$ is the GMM super-vector of a speaker, $m$ is the 
speaker- and channel-independent super-vector (taken from an 
UBM super-vector), $T$ is a rectangular matrix of low rank 
(the total variability matrix) and $w$ is the i-vector, which 
is a random vector with a standard normal distribution
$\mathcal{N}(0,I)$. 
The concept behind this is that a low-dimensional latent vector 
($w$) exists and represents the characteristics of the speaker.
Equation~\ref{eqn:factor-i-vector} can be resolved through 
joint factor analysis, where $w$ represents the factor of the 
total variability matrix ($T$).

In this work we consider a total of 1000 recordings with an
average duration of 22 seconds per recording to train the 
UBM and the $T$ matrix. These recording were provided by
Konecta Group S.A.S.\textregistered. 
The recordings were randomly selected and do not come from the 
same speakers of the KONECTADB.
The implementation of this approach was performed with 
the Kaldi toolkit~\cite{Povey_ASRU2011}. 

\subsubsection{x-vectors}
These are Deep Neural Network (DNN) embedding features, which 
were trained for speaker recognition and verification. 
Table~\ref{tab:x-vector} shows details of the architecture of the
DNN to extract the x-vectors. 

 
\begin{table}[h]
\centering
\caption{Embedding DNN architecture~\cite{xvectors}. x-vectors are extracted in segment6. \textbf{$N$:} Number of training speakers.}
\label{tab:x-vector}
\begin{tabular}{lcccc}
\toprule
\textbf{Layer}         & \textbf{Layer Context} & \textbf{Total Context} & \textbf{Input x output} \\ \hline
\textbf{Frame1}        & ${[}t-2 , t+2{]}$        & 5                      & 120x512                 \\ 
\textbf{Frame2}        & $\{t-2, t, t+2\}$       & 9                      & 1536x512                \\ 
\textbf{Frame3}        & $\{t-3, t, t+3\}$        & 15                     & 1536x512                \\ 
\textbf{Frame4}        & $\{t\}$                  & 15                     & 512x512                 \\ 
\textbf{Frame5}        & $\{t\}$                  & 15                     & 512x1500                \\ 
\textbf{Stats pooling} & ${[}0, T)$               & $T$                      & 1500Tx3000              \\ 
\textbf{Segment6}      & $\{0\}$                  & $T$                       & 3000x512                \\ 
\textbf{Segment7}      & $\{0\}$                  & $T$                      & 512x512                 \\ 
\textbf{Softmax}       & $\{0\}$                  & $T$                       & 512x$N$                  \\ \bottomrule
\end{tabular}
\end{table}
The input features are MFCCs extracted from 24-dimensional filter 
banks with a frame-length of 25ms and a step-size of 10ms. 
The spectrogram is mean-normalized over a sliding window of up 
to 3 seconds. The first five layers operate at a frame level and 
build their context according to the previous layer. 
The stats pooling computes the mean and standard deviation 
from the output of the Frame5 layer using all $T$ frames of 
the signal. The mean and standard deviation are concatenated 
and propagated through the segment-level layers. 
Finally, the softmax layer predicts the speaker. 
The output of the segment6 layer is extracted to create the x-vector.
The implementation of this approach was also performed with 
the Kaldi toolkit~\cite{Povey_ASRU2011}. 

\subsubsection{The I2010PC feature set}
The I2010PC feature set is extracted by following three steps: 
(1) The 38 low-level descriptors show in Table~\ref{table:I2010PC} are 
computed with a step-size of 10ms and a Hanning window with 25ms of length 
(except pitch related features, which are extracted from Gaussian windows 
of 60ms); all instances are smoothed using a moving average filter of 
3 frames. 
(2) Besides the low-level descriptors, 38 first-order regression 
coefficients are included. 
(3) A total of 21 statistical functionals shown in 
Table~\ref{table:I2010PC} are computed per feature vector.
More information and details about how to extract these features 
can be found in~\cite{schuller2010interspeech}.

\begin{table}[h]
\centering
\caption{The low-level descriptors and the functionals.}
\label{table:I2010PC}
\begin{tabular}{l|l}
\toprule
\multicolumn{1}{l|}{\textbf{Descriptors}} & \multicolumn{1}{l}{\textbf{Functionals}} \\ \hline
PCM loudness                               & Position max./min.                        \\
MFCC {[}0-14{]}                            & arith. mean, std. desviation              \\
log Mel Freq. Band {[}0-7{]}               & skewness, kurtosis                        \\
LSP Frequency {[}0-7{]}                    & lin. regression coeff. 1/2                \\
F0                                         & lin. regression error Q/A                 \\
F0 Envelope                                & quartile 1/2/3                            \\
Voicing Prob.                              & quartile range 2-1/3-2/3-1                \\
Jitter local                               & percentile 1/99                           \\
Jitter consec. frame pairs                 & percentile range 99-1                     \\
Shimmer local                              & up-level time 75/90                       \\ \bottomrule
\end{tabular}
\end{table}

\subsubsection{Phonation, articulation and prosody features}
These features are extracted with the DisVoice 
framework~\cite{orozco2018neurospeech},
which was originally developed to model neurological disorders and 
now we want to evaluate its suitability to model emotional speech signals
and customer satisfaction. The source code to extract the features presented 
in this subsection can be found in~\cite{orozco2018neurospeech}.

\textbf{Phonation:}

The phonatory characteristics of a speaker have been typically analyzed 
in terms of features related to perturbation measures such as jitter 
(temporal changes in the fundamental frequency), shimmer 
(amplitude changes in the signal), amplitude perturbation quotient 
(APQ), and pitch perturbation quotient (PPQ). 
APQ and PPQ are long-term perturbation measures of the amplitude and 
the fundamental frequency of the signal, respectively. 
Fuller et al. found that perturbation measures like jitter are a 
reliable indicator of stressor-provoked anxiety~\cite{fuller1992validity}.
Additionally, the results reported in~\cite{li2007stress} show that 
jitter and shimmer are useful to model emotion and stress patterns in 
speech.
The phonation features considered in this work include the first and 
second derivative of the fundamental frequency, jitter, shimmer, APQ,
PPQ, and logarithmic energy. These measures are computed upon 
voiced segments. The global representation per speaker consists of the mean, 
standard deviation, skewness, and kurtosis of the resulting feature vector.

\textbf{Articulation:}
This feature set is inspired in the fact that the transition between 
voiced and unvoiced segments encodes relevant information about 
the capability of the speaker to produce well-articulated utterances.
In previous studies we have shown that this approach is valid for
neurological disorders where speech production is 
affected~\cite{Rafa2016,orozco2018neurospeech}.
Now we want to evaluate its suitability to model emotional speech 
and customer satisfaction. The main hypothesis is that people under
stress due to a bad quality of service or simply due to the need for a
quick solution, are prone to produce more hesitations while speaking.
This phenomenon could be associated to abnormal energy patterns in the
vicinity of the border between voiced and unvoiced sounds, i.e., around
the time when vocal fold vibration starts and/or finishes.

Besides the aforementioned hypothesis, the use of the 
Teager Energy Operator (TEO) is introduced. The authors 
in~\cite{teager1980some} show that the source of speech production 
is not actually a laminar airflow. Instead, it consists of vortex-flow
interactions with the vocal tract boundaries. There is evidence that 
shows these air vortices to be generated during the early opening 
phase and the latter closing phase of the vocal fold, as occurred 
during the transitions between voiced and unvoiced segments~\cite{he2010stress}.
In addition, it is believed that changes in vocal system physiology 
induced by stressful and/or fearful conditions such as muscle tension 
affect the vortex-flow interaction patterns in the vocal
tract~\cite{zhou2001nonlinear}. 
These changes directly affect the spectrum of the speech signal. 
Thus, feature sensitive to the presence of these additional vortices 
between the transitions could help to detect the emotional state 
of the speaker.
In this case the main assumption is that if the speaker's speech is 
altered due to changes in the emotional state, then such changes 
will produce abnormal articulation patterns during speech production. 

The complete articulation feature set includes 
the first two formants with their first and second derivatives 
extracted from the voiced segments. The energy in the 
transitions between voiced and unvoiced segments which is computed
and distributed according to the Bark bands, and MFCCs with their 
first and second derivatives are also measured in the aforementioned 
transitions.

\textbf{Prosody:} 
Two types of prosodic features can be found in the literature: basic and 
compound. Basic prosody attributes include loudness, pitch, voice quality,
duration, speaking rate, and pauses. 
Variations of these measures over time constitute the compound 
prosodic attributes, which are intonation, accentuation, prosodic 
phrases, rhythm, and hesitation~\cite{kompe1997prosody}. 
Human beings typically use prosody features to identify emotions 
present in daily conversations~\cite{rao2013emotion}. 
For instance, in active emotions like anger, pitch and energy 
are high while they are relatively low in passive emotions like 
sadness. This paper considers prosodic features based on duration, 
fundamental frequency, and energy to model emotional patterns 
in speech. 
Several statistical functionals are computed per feature vector including 
mean value, standard deviation, skewness, kurtosis, maximum, and 
minimum. 
Details of the methods and algorithms to extract these 
features can be found in~\cite{orozco2018neurospeech}.

\subsection{\label{sec:classification}Classification and Evaluation}
The classification process is performed with a soft-margin 
SVM with a Gaussian kernel. 
This classifier has two parameters, complexity $C$ and the 
kernel bandwidth $\gamma$. 
These two parameters are optimized in a grid-search up to powers
of ten where $C \in [10^{-3}, \ldots, 10^{4}]$ and
$\gamma \in [10^{-6}, \ldots, 10^{3}]$. 
To avoid biased or optimistic results, a nested cross-validation
strategy is followed~\cite{parvandeh2020consensus}. 
We considered 5 folds for outer and also for inner cross-validation.
Both scenarios, speaker-independent and speaker-dependent, are 
considered in the experiments about emotion classification.

\section{Experiments and results~\label{sec:results}}
Different experiments are performed with the different datasets
considered in this study. Experiments with IEMOCAP, EMODB, and RAVDESS
are all multi-class, i.e., all emotions included in these corpora are
considered. The experiments with the KONECTADB corpus
are bi-class, where the aim is to discriminate between satisfied vs.
dissatisfied customers.
The results are reported in terms of Unweighted Average Recall (UAR) 
and Accuracy (ACC). 
Optimal hyper-parameters found for each classification experiment
are also reported to allow direct comparisons in future studies.
The symbol $+$ in the feature representation 
means fusion with other feature sets, while art, pro, and pho are
articulation, prosody, and phonation, respectively.

\subsection{Experiments with IEMOCAP}
Table~\ref{tab:performance-iemocap} shows the performance of the
classifier on the IEMOCAP database. 
Note that, for both speaker-independent and speaker-dependent
scenarios, the best results are obtained with the I2010PC feature 
set with UARs of 58.9\% and 68.1\%, respectively.
The second best approach is the one based on x-vectors which yields  
UARs of 57.6\% and 65.4\%, respectively. The third best 
model is the combination `art+pro+pho' which achieved 
57.0\% and 61.3\% of UAR, respectively.
It is important to highlight that the third best model is very close 
to I2010PC in the speaker-independent scenario, which is the most 
realistic and unbiased one. Additionally, note that the
proposed approach uses much less features in comparison to the 
I2010PC feature set which extracts a total of 1582 features. 

\setlength{\tabcolsep}{1pt}
\begin{table}[ht!]
\centering
\caption{\label{tab:performance-iemocap}Results of multi-class classification of emotions of the IEMOCAP database. \textbf{ACC}: Accuracy. \textbf{UAR}: Unweighted Average Recall. The performance metrics are given in [\%]}
\begin{tabular}{lcccccccc}
\toprule
\multirow{2}{*}{\textbf{Feature set}} & \multicolumn{4}{c}{\textbf{Speaker independent}}                  & \multicolumn{4}{c} {\textbf{Speaker dependent}}                \\
                                    & $C$ & \phantom{s}$\gamma$ & \textbf{\phantom{a}UAR\phantom{a}} & \textbf{\phantom{a}ACC\phantom{a}} & $C$ & \phantom{s}$\gamma$ & \textbf{\phantom{a}UAR\phantom{a}} & \textbf{\phantom{a}ACC\phantom{a}} \\ \hline
I2010PC                             & 10          & \phantom{s}0.001                           & 58.9     & 57.4        & 1          & \phantom{s}0.001                           & 68.1      & 67.2       \\
i-vector                            & 1          & \phantom{s}0.01                          & 53.5    & 51.1         & 1          & \phantom{s}0.01                          & 60.2   & 57.5          \\
x-vector                            & 1          & \phantom{s}0.001                           & 57.6    & 55.9          & 1          & \phantom{s}0.001                           & 65.4 & 63.9             \\
articulation                             & 1          & \phantom{s}0.001                           & 54.0      & 51.9        & 1          & \phantom{s}0.001                           & 59.5      & 57.2       \\
prosody                             & 1          & \phantom{s}0.0001                           & 44.5      & 41.2        & 100          & \phantom{s}0.0001                           & 48.1      & 45.2      \\
phonation                             & 1          & \phantom{s}0.001                           & 46.0      & 43.0        & 100          & \phantom{s}0.001                           & 48.5      & 45.4      \\
art+pro                             & 1          & \phantom{s}0.001                           & 56.1      & 54.2       & 1          & \phantom{s}0.001                           & 60.7      & 58.5      \\
art+pho                             & 1          & \phantom{s}0.001                           & 56.0      & 53.6        & 1          & \phantom{s}0.001                           & 59.9      & 57.6      \\
pro+pho                             &  1         & \phantom{s}1e-5                           & 47.5      & 44.9        & 10          & \phantom{s}0.0001                           & 50.4      & 47.6      \\
art+pro-pho                             & 1          & \phantom{s}0.001                           & 57.0      & 54.7        & 1          & \phantom{s}0.001                           & 61.3      & 59.3      \\
\bottomrule

\end{tabular}
\end{table}

\subsection{Experiments with EMODB}
The results obtained with the EMODB corpus are shown in 
Table~\ref{tab:performance-emodb}. 
In the speaker-independent scenario the best UAR is 74\% which is
obtained with the I2010PC feature set. The second and the third
best models are `art+pro' and `art+pho' with UARs of 
64\% and 61.5\%, respectively. In the speaker-dependent scenario
the best three models are x-vectors, I2010PC, and i-vectors,
respectively.
These results confirm the capability of the x-vectors to model
speaker characteristics~\cite{raj2019probing}. 

Although some studies reviewed in Section~\ref{sec:state-of-the-art}
reported better results, those were optimistic because they 
considered the same set to train and evaluate the classifier. 
For example, the authors in~\cite{eyben2015geneva,wang2015speech} 
used an SVM to do the classification and reported accuracies above 
79\% on the speaker-independent scenario. However, these 
results were achieved on the same set that was used to optimize the
classifier. 
Instead, we followed a nested cross-validation strategy, where
the test set is unseen by the optimal parameters, leading to
more realistic and unbiased results.

\setlength{\tabcolsep}{1pt}
\begin{table}[ht!]
\centering
\caption{\label{tab:performance-emodb}Results of multi-class classification of emotions of the EMODB database. \textbf{ACC}: Accuracy. \textbf{UAR}: Unweighted Average Recall. The performance metrics are given in [\%]}
\begin{tabular}{lcccccccc}
\toprule
\multirow{2}{*}{\textbf{Feature set}} & \multicolumn{4}{c}{\textbf{Speaker independent}}                  & \multicolumn{4}{c}{\textbf{Speaker dependent}}                \\
                                    & $C$ & \phantom{s}$\gamma$ & \textbf{\phantom{a}UAR\phantom{a}} & \textbf{\phantom{a}ACC\phantom{a}} & $C$ & \phantom{s}$\gamma$ & \textbf{\phantom{a}UAR\phantom{a}} & \textbf{\phantom{a}ACC\phantom{a}} \\ \hline
I2010PC                             & 10          & \phantom{s}0.0001                           & 74.0      & 73.3        & 10          & \phantom{s}0.001                           & 84.6      & 85.6       \\
i-vector                            & 10          & \phantom{s}0.01                          & 59.6    & 60.7         & 10          & \phantom{s}0.01                          & 73.6   & 74.2          \\
x-vector                            & 10          & \phantom{s}0.0001                           & 60.9    & 61.4          & 10          & \phantom{s}0.0001                           & 86.0 & 86.0             \\
articulation                             & 1          & \phantom{s}0.001                           & 58.9      & 62.7        & 10          & \phantom{s}0.001                           & 69.3      & 71.2       \\
prosody                             & 10          & \phantom{s}0.01                           & 56.4      & 58.8        & 10          & \phantom{s}0.01                           & 54.7      & 56.9      \\
phonation                             & 1000          & \phantom{s}0.001                           & 50.3      & 52.2        & 10          & \phantom{s}0.01                           & 52.5      & 53.9      \\
art+pro                             & 10          & \phantom{s}0.001                           & 64.0      & 67.6       & 1          & \phantom{s}0.001                           & 59.3      & 61.2      \\
art+pho                             & 10          & \phantom{s}0.0001                           & 61.5      & 65.0        & 1          & \phantom{s}0.001                           & 60.7      & 63.5      \\
pro+pho                             & 1          & \phantom{s}0.001                           & 53.9      & 55.1        & 1          & \phantom{s}0.01                           & 58.9      & 62.7      \\
art+pro+pho                             & 10          & \phantom{s}0.0001                           & 55.1      & 57.7        & 1          & \phantom{s}0.001                           & 50.5      & 50.7      \\
\bottomrule
\end{tabular}
\end{table}

\subsection{Experiments with RAVDESS}
Table~\ref{tab:performance-ravdess} shows the results obtained with 
the RAVDESS database. 
Similar to the results obtained in the previous experiments,
in the speaker-independent scenario the best three models are 
I2010PC, x-vectors, and `art+pro+pho', with UARs of
58.7\%, 58.6\%, and 47.1\%, respectively.
In the speaker-dependent scenario the best model is the one
based on x-vectors with an UAR of 83.4\%, while the second and
third best models are I2010PC and i-vectors with UARs of 70.9\% and
64.6\%, respectively.

There are several works, most of them based on deep learning, 
that achieve accuracies of up to 70.0\% in the speaker-independent
scenario using the RAVDESS database~\cite{issa2020speech,sajjad2020clustering,jalal2019learning}. 
However, it is not possible to know whether those results are 
based on unseen data. 
In the speaker-dependent scenario, x-vector achieved 
better performance (83.4\% ACC) than those obtained by the 
authors in~\cite{sajjad2020clustering} (82.41\% ACC), where 
a more complex model was proposed.

\setlength{\tabcolsep}{1pt}
\begin{table}[ht!]
\centering
\caption{\label{tab:performance-ravdess}Results of multi-class classification of emotions of the RAVDESS database. \textbf{ACC}: Accuracy. \textbf{UAR}: Unweighted Average Recall. The performance metrics are given in [\%]}
\begin{tabular}{lcccccccc}
\toprule
\multirow{2}{*}{\textbf{Feature set}} & \multicolumn{4}{c}{\textbf{Speaker independent}}                  & \multicolumn{4}{c}{\textbf{Speaker dependent}}                \\
                                    & $C$ & \phantom{s}$\gamma$ & \textbf{\phantom{a}UAR\phantom{a}} & \textbf{\phantom{a}ACC\phantom{a}} & $C$ & \phantom{s}$\gamma$ & \textbf{\phantom{a}UAR\phantom{a}} & \textbf{\phantom{a}ACC\phantom{a}} \\ \hline
I2010PC                             & 10          & \phantom{s}1e-5                          & 58.7      & 57.6        & 10          & \phantom{s}1e-4                           & 70.9      & 70.5       \\
i-vector                            & 1          & \phantom{s}0.01                          & 44.7    & 45.3         & 10          & \phantom{s}0.01                          & 64.6   & 64.1          \\
x-vector                            & 10          & \phantom{s}0.001                           & 58.6    & 58.7          & 10          & \phantom{s}0.001                           & 83.4 & 83.5\\
articulation                             & 10          & \phantom{s}0.001                           & 43.6      & 44.4        & 1          & \phantom{s}0.001                           & 46.4      & 46.6      \\
prosody                             & 1          & \phantom{s}0.01                           & 37.0      & 36.5        & 10          & \phantom{s}0.01                           & 39.1      & 38.8      \\
phonation                             & 1e5          & \phantom{s}1e-4                           & 34.1      & 33.5        & 100          & \phantom{s}0.001                           & 37.8      & 36.8      \\
art+pro                             & 10          & \phantom{s}0.001                           & 46.9      & 48.1       & 10          & \phantom{s}0.001                           & 53.8      & 54.5      \\
art+pho                             & 1          & \phantom{s}0.001                           & 46.9      & 47.2        & 10          & \phantom{s}0.001                           & 46.9      & 47.4      \\
pro+pho                             & 10          & \phantom{s}0.01                           & 39.0      & 39.0        & 10          & \phantom{s}0.001                           & 40.0      & 39.9      \\
art+pro+pho                             & 10          & \phantom{s}0.001                           & 47.1      & 47.8        & 10          & \phantom{s}0.001                           & 53.9      & 54.4      \\
\bottomrule
\end{tabular}
\end{table}

\subsection{KONECTADB experiment\label{sec:konecta-experiment}}

\begin{figure*}[!h]
\centering
   \centering
   \includegraphics[width=\textwidth]{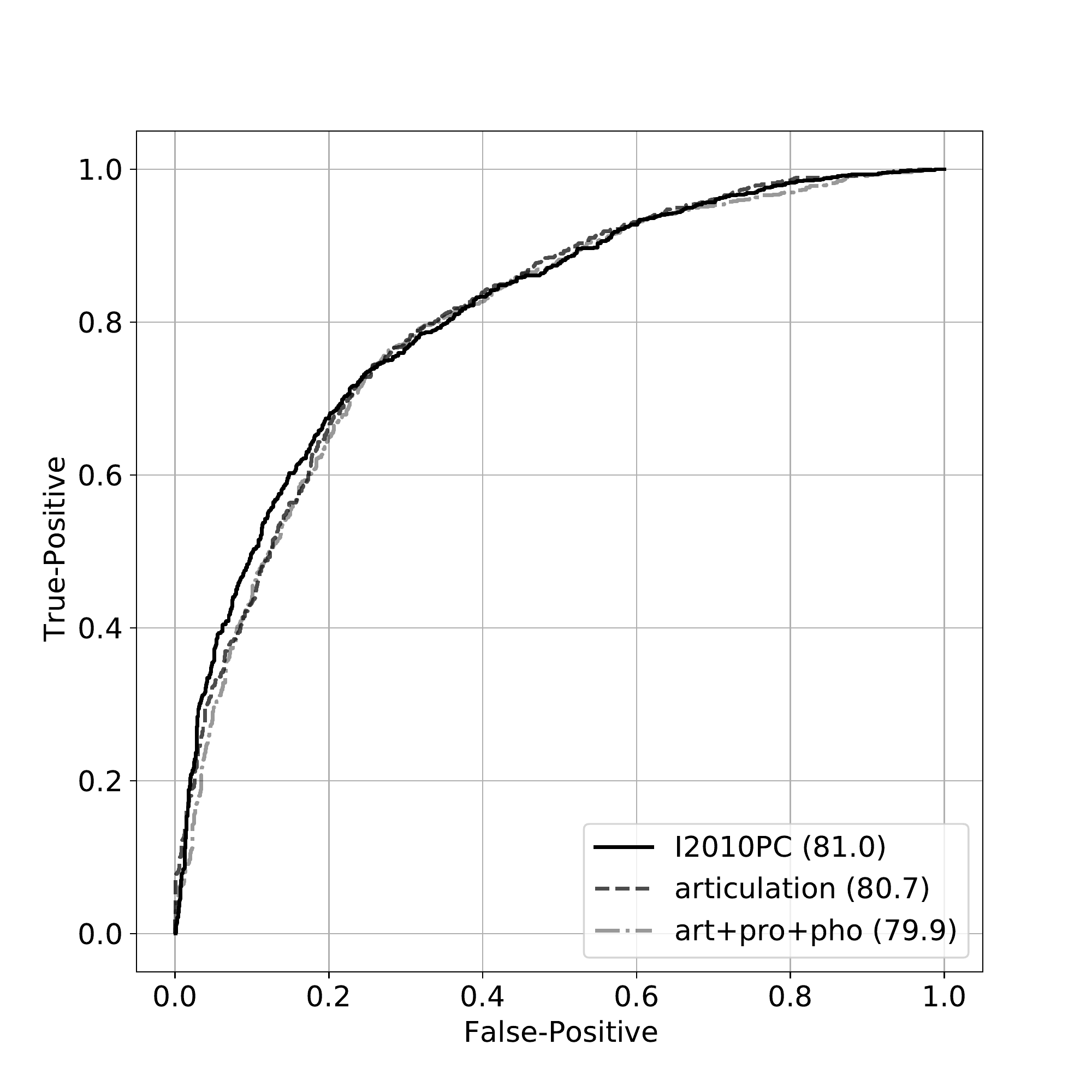}
\hfill
\caption{ROC curves obtained with I2010PC, articulation, and art+pro+pho feature sets in the classification of satisfied vs. dissatisfied customers.}
\label{fig:roc-curve}
\end{figure*}

All recordings in the KONECTADB corpus were collected from different 
customers and only the speaker-independent scenario was addressed. Table~\ref{tab:performance-konecta} shows the results obtained when
classifying between satisfied and dissatisfied customers. 
Note that the best performance is achieved with the `articulation' feature 
set with an accuracy of 74.3\%. The second best ACC is obtained with the 
`art+pro+pho' model (73.8\%), and the third best result is observed with
the I2010PC feature set (73.6\%). It is important to highlight that in these
experiments, which are based on recordings collected under non-controlled
acoustic conditions and considering non-acted opinions of the received service,
speaker models -based approaches do not show good performances, which
is opposite to what we observed in some of the previous results where
the x-vector approach seemed to be competitive. 
Besides numerical results, ROC curves are included in
Figure~\ref{fig:roc-curve} with the aim 
to show the results more compactly. Note that the  
overall performance of the top three methods is similar.

\begin{table}[ht!]
\centering
\caption{\label{tab:performance-konecta}Results of the customer satisfaction classification (satisfied vs. dissatisfied) in the KONECTADB database. \textbf{ACC}: Accuracy. \textbf{UAR}: Unweighted Average Recall. \textbf{SEN}: Sensitivity. \textbf{SPE}: specificity.
The performance metrics are given in [\%]
}
\begin{tabular}{lcccccc}
\toprule
\textbf{Feature set}  & $C$  & $\gamma$  & \textbf{\phantom{a}UAR\phantom{a}}  & \textbf{\phantom{a}ACC\phantom{a}}  & \textbf{\phantom{a}SEN\phantom{a}}  & \textbf{\phantom{a}SPE\phantom{a}}  \\ \hline
I2010PC    & 10   & 1e-4               & 72.8 & 73.6  & 62.8 & 82.9 \\
i-vector     & 100    & 0.001      & 59.6 & 61.2  & 60.8 & 62.8 \\ 
x-vector     & 10 & 1e-4          & 67.3 & 67.6 & 62.0 & 72.5 \\
prosody      & 1    & 1e-4               & 66.0 & 66.4 & 60.7 & 71.2 \\
articulation & 0.1  & 0.001              & 74.2 & 74.3 & 72.4 & 76.0 \\
phonation    & 100  & 0.001              & 67.0 & 67.4 & 60.6 & 73.4 \\
art+pro      & 0.1  & 0.001              & 68.9 & 69.7 & 57.7 & 80.1 \\
art+pho      & 0.1  & 0.001              & 68.8 & 69.0 & 66.2 & 71.4 \\
pro+pho      & 2000 & 1e-5             & 64.1 & 63.2 & 77.0 & 51.3 \\
art+pro+pho  & 1    & 0.01             & 73.6 & 73.8 & 69.9 & 77.3 \\
\bottomrule
\end{tabular}
\end{table}

The results presented above likely indicate that the introduced approach 
is more robust against non-controlled acoustic conditions, which are realistic
and may include channel distortions, microphone imperfections, and highly 
variable acoustic conditions.

\section{Conclusions and discussion~\label{sec:conclusions}}
This work evaluates different ``standard' feature sets typically used 
to classify emotions in speech, and also presents a novel approach based on modeling 
phonation, articulation and prosody aspects
that may vary when the emotional state of the speaker is changed.
The methods are also evaluated in the problem of classifying customer 
satisfaction based on speech recordings where customers give their 
opinion about the received service.
The databases with emotional speech recordings are those typically used
in the literature, which consider acted emotions, limited number of speakers, and
relatively controlled acoustic conditions. Conversely, the database recorded
in the call-center was collected without any control over the communication
channel or microphone, and considers recordings of real opinions (i.e., non-acted)
of the customers about the received service.

The results show that the I2010PC features achieved the best results 
when classifying emotional speech. In this scenario, features included in 
our proposed approach were in the top-three of the best feature sets.
For the case of customer satisfaction evaluation, our approach based on
articulation features yields the best results, followed by the combination of 
articulation, phonation, and prosody features. The third best feature set
is I2010PC.
Regarding the speaker models, x-vectors outperformed the rest of approaches in two
of the three databases where the speaker-dependent scenario was considered.
This was expected since this method has shown excellent results in modeling
speaker-specific information.

We believe that our approach is highly competitive considering for instance, 
the reduced number of features extracted with the articulation approach (488)
compared to the 1584-dimensional feature vector extracted with the I2010PC toolkit.
Additionally, the results when considering non-controlled acoustic conditions 
and non-acted opinions about a received service make us to think that 
the proposed approach is more convenient for industrial applications.

\section*{Acknowledgements}
This work received funding from CODI from the University of Antioquia grant \# PRG2020-34068.

\bibliography{mybibfile}

\begin{thebibliography}{10}
\expandafter\ifx\csname url\endcsname\relax
  \def\url#1{\texttt{#1}}\fi
\expandafter\ifx\csname urlprefix\endcsname\relax\def\urlprefix{URL }\fi
\expandafter\ifx\csname href\endcsname\relax
  \def\href#1#2{#2} \def\path#1{#1}\fi

\bibitem{zweig2006automated}
G.~Zweig, O.~Siohan, G.~Saon, B.~Ramabhadran, D.~Povey, L.~Mangu, B.~Kingsbury,
  Automated quality monitoring in the call center with {ASR} and maximum
  entropy, in: Proceedings of the IEEE International Conference on Acoustics,
  Speech and Signal Processing, 2006, pp. 589--592.

\bibitem{mishne2005automatic}
G.~Mishne, D.~Carmel, R.~Hoory, A.~Roytman, A.~Soffer, Automatic analysis of
  call-center conversations, in: Proceedings of the ACM International
  Conference on Information and Knowledge Management, 2005, pp. 453--459.

\bibitem{paralinguistics2013}
B.~Schuller, S.~Steidl, A.~Batliner, F.~Burkhardt, L.~Devillers, C.~M{\"u}Ller,
  S.~Narayanan, Paralinguistics in speech and language --state-of-the-art and
  the challenge, Computer Speech \& Language 27 (2013) 4--39.

\bibitem{eyben2015geneva}
F.~Eyben, K.~Scherer, B.~Schuller, J.~Sundberg, E.~Andr{\'e}, C.~Busso,
  L.~Devillers, J.~Epps, P.~Laukka, S.~Narayanan, et~al., The {G}eneva
  minimalistic acoustic parameter set ({GeMAPS}) for voice research and
  affective computing, IEEE Transactions on Affective Computing 7 (2015)
  190--202.

\bibitem{guo2019exploration}
L.~Guo, L.~Wang, J.~Dang, Z.~Liu, H.~Guan, Exploration of complementary
  features for speech emotion recognition based on kernel extreme learning
  machine, IEEE Access 7 (2019) 75798--75809.

\bibitem{xia2012using}
R.~Xia, Y.~Liu, Using i-vector space model for emotion recognition, in:
  Proceedings of the Annual Conference of the International Speech
  Communication Association, 2012, pp. 2230--2233.

\bibitem{jalal2019learning}
M.~Jalal, E.~Loweimi, R.~Moore, T.~Hain, Learning temporal clusters using
  capsule routing for speech emotion recognition., in: Proceedings of the
  Annual Conference of the International Speech Communication Association,
  2019, pp. 1701--1705.

\bibitem{khaki2015continuous}
H.~Khaki, E.~Erzin, Continuous emotion tracking using total variability space,
  in: Proceedings of the Annual Conference of the International Speech
  Communication Association, 2015, pp. 1299--1303.

\bibitem{pappagari2020x}
R.~Pappagari, T.~Wang, J.~Villalba, N.~Chen, N.~Dehak, x-vectors meet emotions:
  A study on dependencies between emotion and speaker recognition, in:
  Proceedings of the IEEE International Conference on Acoustics, Speech and
  Signal Processing, 2020, pp. 7169--7173.

\bibitem{orozco2018neurospeech}
J.~Orozco-Arroyave, J.~V{\'a}squez-Correa, J.~Vargas-Bonilla, R.~Arora,
  N.~Dehak, P.~Nidadavolu, H.~Christensen, F.~Rudzicz, M.~Yancheva, H.~Chinaei,
  et~al., Neurospeech: An open-source software for parkinson's speech analysis,
  Digital Signal Processing 77 (2018) 207--221.

\bibitem{wang2015speech}
K.~Wang, N.~An, B.~N. Li, Y.~Zhang, L.~Li, Speech emotion recognition using
  fourier parameters, IEEE Transactions on Affective Computing 6 (2015) 69--75.

\bibitem{semwal2017}
N.~Semwal, A.~Kumar, S.~Narayanan, Automatic speech emotion detection system
  using multi-domain acoustic feature selection and classification models, in:
  Proceedings of the IEEE International Conference on Identity, Security and
  Behavior Analysis, 2017, pp. 1--6.

\bibitem{busso2004analysis}
C.~Busso, Z.~Deng, S.~Yildirim, M.~Bulut, C.~Lee, A.~Kazemzadeh, S.~Lee,
  U.~Neumann, S.~Narayanan, Analysis of emotion recognition using facial
  expressions, speech and multimodal information, in: Proceedings of the
  International Conference on Multimodal Interfaces, 2004, pp. 205--211.

\bibitem{iemocap}
C.~Busso, M.~Bulut, C.~Lee, A.~Kazemzadeh, E.~Mower, S.~Kim, J.~Chang, S.~Lee,
  S.~Narayanan, {IEMOCAP}: Interactive {E}motional {D}yadic {M}otion {C}apture
  {D}atabase, Language Resources and Evaluation 42~(4) (2008) 335.

\bibitem{schuller2010interspeech}
B.~Schuller, S.~Steidl, A.~Batliner, F.~Burkhardt, L.~Devillers, C.~M{\"u}ller,
  S.~Narayanan, The {INTERSPEECH} 2010 paralinguistic challenge, in:
  Proceedings of the Annual Conference of the International Speech
  Communication Association, 2010, pp. 2794--2797.

\bibitem{fayek2017evaluating}
H.~Fayek, M.~Lech, L.~Cavedon, Evaluating deep learning architectures for
  speech emotion recognition, Neural Networks 92 (2017) 60--68.

\bibitem{issa2020speech}
D.~Issa, M.~Demirci, A.~Yazici, Speech emotion recognition with deep
  convolutional neural networks, Biomedical Signal Processing and Control 59.

\bibitem{sajjad2020clustering}
M.~Sajjad, S.~Kwon, et~al., Clustering-based speech emotion recognition by
  incorporating learned features and deep bilstm, IEEE Access 8 (2020)
  79861--79875.

\bibitem{livingstone2018ryerson}
S.~R. Livingstone, F.~A. Russo, The {R}yerson {A}udio-{V}isual {D}atabase of
  {E}motional {S}peech and {S}ong ({RAVDESS}): A dynamic, multimodal set of
  facial and vocal expressions in north american english, PloS one 13.

\bibitem{burkhardt2005database}
F.~Burkhardt, A.~Paeschke, M.~Rolfes, W.~F. Sendlmeier, B.~Weiss, A database of
  german emotional speech, in: Ninth european conference on speech
  communication and technology, 2005.

\bibitem{Povey_ASRU2011}
D.~Povey, A.~Ghoshal, G.~Boulianne, L.~Burget, O.~Glembek, N.~Goel,
  M.~Hannemann, P.~Motlicek, Y.~Qian, P.~Schwarz, J.~Silovsky, G.~Stemmer,
  K.~Vesely, The {K}aldi {S}peech {R}ecognition {T}oolkit, in: Proceedings of
  the IEEE Workshop on Automatic Speech Recognition and Understanding, 2011.

\bibitem{campbell2006support}
W.~Campbell, D.~Sturim, D.~Reynolds, {S}upport {V}ector {M}achines using {GMM}
  supervectors for speaker verification, IEEE Signal Processing Letters 13
  (2006) 308--311.

\bibitem{campbell2006svm}
W.~Campbell, D.~Sturim, D.~Reynolds, A.~Solomonoff, {SVM} based speaker
  verification using a {GMM} supervector kernel and {NAP} variability
  compensation, in: Proceedings of the IEEE International Conference on
  Acoustics, Speech and Signal processing, 2006, pp. 97--100.

\bibitem{you2008svm}
C.~You, K.~Lee, H.~Li, An {SVM} kernel with {GMM}-supervector based on the
  {B}hattacharyya distance for speaker recognition, IEEE Signal Processing
  Letters 16~(1) (2008) 49--52.

\bibitem{dehak2009support}
N.~Dehak, R.~Dehak, P.~Kenny, N.~Br{\"u}mmer, P.~Ouellet, P.~Dumouchel, Support
  {V}ector {M}achines versus fast scoring in the low-dimensional total
  variability space for speaker verification, in: Proceedings of the Annual
  Conference of the International Speech Communication Association, 2009, pp.
  1559--1562.

\bibitem{xvectors}
D.~Snyder, D.~Garcia-Romero, G.~Sell, D.~Povey, S.~Khudanpur, x-vectors: Robust
  {DNN} embeddings for speaker recognition, in: Proceedings of the IEEE
  International Conference on Acoustics, Speech and Signal Processing, 2018,
  pp. 5329--5333.

\bibitem{fuller1992validity}
B.~Fuller, Y.~Horii, D.~Conner, Validity and reliability of nonverbal voice
  measures as indicators of stressor-provoked anxiety, Research in Nursing \&
  Health 15 (1992) 379--389.

\bibitem{li2007stress}
X.~Li, J.~Tao, M.~Johnson, J.~Soltis, A.~Savage, K.~Leong, J.~Newman, Stress
  and emotion classification using jitter and shimmer features, in: Proceedings
  of the IEEE International Conference on Acoustics, Speech and Signal
  Processing, 2007, pp. 1081--1084.

\bibitem{Rafa2016}
J.~Orozco-Arroyave, Analysis of speech of people with parkinson's disease,
  Vol.~41, Logos-Verlag, 2016.

\bibitem{teager1980some}
H.~Teager, Some observations on oral air flow during phonation, IEEE
  Transactions on Acoustics, Speech, and Signal Processing 28 (1980) 599--601.

\bibitem{he2010stress}
L.~He, Stress and emotion recognition in natural speech in the work and family
  environments, PhD, Rmit University (2010) 1--218.

\bibitem{zhou2001nonlinear}
G.~Zhou, J.~Hansen, J.~Kaiser, Nonlinear feature based classification of speech
  under stress, IEEE Transactions on Speech and Audio Processing 9 (2001)
  201--216.

\bibitem{kompe1997prosody}
R.~Kompe, R.~Kompe, Prosody in speech understanding systems, Vol. 1307,
  Springer, 1997.

\bibitem{rao2013emotion}
K.~Rao, S.~Koolagudi, R.~Vempada, Emotion recognition from speech using global
  and local prosodic features, International Journal of Speech Technology 16
  (2013) 143--160.

\bibitem{parvandeh2020consensus}
S.~Parvandeh, H.~Yeh, M.~Paulus, B.~McKinney, Consensus features nested
  cross-validation, Bioinformatics 36 (2020) 3093--3098.

\bibitem{raj2019probing}
D.~Raj, D.~Snyder, D.~Povey, S.~Khudanpur, Probing the information encoded in
  x-vectors, in: Proceedings of the IEEE Automatic Speech Recognition and
  Understanding Workshop, 2019, pp. 726--733.

\end{thebibliography}

\end{document}